\documentstyle[bbm,amsfonts,epsfig,12pt,here]{article}

\newcommand {\eqref} [1] {(\ref {#1})}
\newcommand {\slsh} [1] {\not{\hbox{\kern-2pt${#1}$}}}

\def\drawbox#1#2{\hrule height#2pt
         \hbox{\vrule width#2pt height#1pt \kern#1pt
               \vrule width#2pt}
               \hrule height#2pt}

\def\Asym#1#2{\vcenter{\vbox{\drawbox{#1}{#2}
               \kern-#2pt       
               \drawbox{#1}{#2}}}}

\newcommand {\beq} {\begin{equation}}
\newcommand {\eeq} {\end{equation}}
  \newcommand {\ber}{\begin{eqnarray*}}
  \newcommand {\eer} {\end{eqnarray*}}
\newcommand {\bea}{\begin{eqnarray}}
  \newcommand {\eea} {\end{eqnarray}}

\newcommand{\None}{${\cal N}=1\ $}

\begin{document}
\begin{titlepage}
\begin{flushright}{CERN-PH-TH/2004-248

SWAT-421

FTPI-MINN-04/33, UMN-TH-2320/04

December 17, 2004}

\end{flushright}
\vskip 1cm

\centerline{{\Large \bf Refining the Proof of Planar Equivalence}}
\vskip 1cm
\centerline{\large A. Armoni ${}^{a,b}$, M. Shifman ${}^{c}$, G. 
Veneziano ${}^{a,d}$}
\vskip 0.1cm

\vskip 0.5cm
\centerline{${}^a$ Department of Physics, Theory Division, CERN}
\centerline{CH-1211 Geneva 23, Switzerland}
\vskip 0.5cm
\centerline{${}^b$ Department of Physics, University of Wales Swansea,}
\centerline{Singleton Park, Swansea, SA2 8PP, UK}
\vskip 0.5cm
\centerline{${}^c$ William I. Fine Theoretical Physics Institute, 
University
of Minnesota,}
\centerline{Minneapolis, MN 55455, USA}
\vskip 0.5cm
\centerline{${}^d$ Coll\`ege de France, 11 place M. Berthelot, 75005 Paris, France}
\vskip 1cm

\begin{abstract}

We outline a full non-perturbative  proof of planar (large-$N$) equivalence
between  bosonic correlators in a theory with Majorana fermions in the adjoint representation and one with Dirac fermions in the two--index (anti)symmetric representation. In a particular case (one flavor), this reduces to our previous
result --- planar equivalence between super-Yang--Mills theory and a 
non-supersymmetric ``orientifold field theory.'' 
The latter theory becomes one-flavor massless QCD at $N=3$.
 
 \end{abstract}

\end{titlepage}

Recently, we have argued \cite{Armoni:2003gp}
that a bosonic sector of \None super-Yang--Mills (SYM) theory 
is equivalent,  in the large-$N$ planar limit, to a
corresponding  sector of a non-supersymmetric  Yang--Mills  theory with
a Dirac fermion in the two--index antisymmetric or symmetric
representation.  We will refer to these  as the parent  and
 daughter theories,  respectively,  all being endowed with the same gauge group, 
SU$(N)$. The daughter theories represent orientifold
projections of the parent one, as first discussed in Ref.\cite{Sag}. As we shall see, all our results apply equally well to the
antisymmetric  (orienti-A), and to the  symmetric  (orienti-S) case.
For a detailed review see Ref.~\cite{Armoni:2004uu}.

For the orienti-A case the daughter theory reduces,
at $N=3$, to one-flavor massless QCD. Thus, as
an intriguing consequence of  planar equivalence, one can copy,  within an ${\cal O}(1/N)$ error, non-perturbative
quantities from SYM theory to the corresponding ones in one-flavor 
massless QCD \cite{Armoni:2003fb}. In particular,  
in \cite{Armoni:2003yv}
we have obtained a  very encouraging  value for the quark condensate.
Orientifold planar equivalence has further possible applications, both in
phenomenology \cite{Sannino:2004qp} and in string theory 
\cite{DiVecchia:2004ev}.

In Refs.~\cite{Armoni:2003gp,Armoni:2004uu} we provided 
a perturbative proof of the planar equivalence and   outlined
a non-perturbative extension of it. In this paper we present a detailed
analysis of non-perturbative  planar equivalence (including theories
with $N_f$ flavors, $N_f>1$), with  emphasis on the assumptions made. In our view, this completes the non-perturbative proof, under very mild assumptions.

The basic idea behind our proof is the comparison
of   generating functionals of   appropriate   gauge-invariant  correlators in the
parent and daughter theories by,  first, integrating out their respective fermions in a fixed gauge background --- a feature which could not be implemented
for the orbifold projection \footnote{Planar equivalence for
``orbifold filed theories'' was
conjectured by M. Strassler in Ref. \cite{Strassler:2001fs}. Orbifold
theories {\em always} contain a product of gauge factors.} --- and,   then, averaging over the gauge field itself.
In Refs.~\cite{Armoni:2003gp,Armoni:2004uu} the main emphasis
was on the first step. Here we mainly focus on the second.

Let us define, for a generic Dirac fermion in the representation $r$, the generating functional,
\beq
\label{defW}
e^{-{\cal W}_r(J_{\rm YM},\,J_\Psi)} =  \int \, DA_\mu \, D\Psi\,  D\bar{\Psi} \, 
e^{ -S_{\rm YM} [A,J_{\rm YM} ]}\, 
\exp \left\{ \bar{\Psi} \left (i  \not\! \partial + \not\!\!  A ^a \, T^a _r + J_\Psi \right){\Psi}\right\}\,,
\eeq
where
$S_{\rm YM}$ is the Yang--Mills action,  $J_{\rm YM}$ is a source
which can couple to any gauge-invariant operator
built from gauge fields and covariant derivatives, and  the quark 
(color-singlet) source
$J_\Psi $ can contain Lorentz $\gamma$ matrices. A mass term is a particular
case of such  quark source. We will always assume that a small 
fermion mass term is introduced for infrared regularization.
It can be set to zero at the very end.
The generating functional 
${\cal W}_r(J_{\rm YM},\,J_\Psi)$
is written in (\ref{defW}), for definiteness, in  Euclidean space.
This is not crucial: one can carry out all our derivations in Minkowski space as well.

After the fermions are integrated out we arrive at
\beq 
\label{fint}
e^{-{\cal W}_r(J_{\rm YM},\,J_\Psi)}= \int \, DA_\mu\,
 e^{ -S_{\rm YM}[A,J_{\rm YM} ] + \Gamma_r [A, J_\Psi]}\,,
\eeq
where 
\beq
\label{defGa}
 \Gamma _r [A, J_\Psi]=\log \, \det \left( i\not\! \partial + \not\!\!  A ^a
\, T^a _r + J_\Psi\right) \, . 
\\
 \eeq
For what follows it is convenient to 
 write the effective action $\Gamma _r [A, J_\Psi]$ in the world-line
formalism, see  \cite{Casalbuoni,Brink:1976uf,Strassler:1992zr,D'Hoker:1995ax}, as an
integral over (super-)Wilson loops, namely\footnote{Strictly speaking Eq.\eqref{wlineint} is only valid for space-time
 independent currents proportional to $1$ or $\gamma_5$. The extension
 to non-constant currents can be found in \cite{D'Hoker:1995ax}
 for those $\gamma$-matrix structures
and do not affect our considerations below. As discussed below and at
the end of the paper, we also expect the same to be true for other
$\gamma$-matrix structures (see \cite{D'Hoker:1995bj}) provided suitable identifications are made for
the currents in the various theories we consider.}

\bea
\label{wlineint}
 \Gamma _r [A, J_\Psi] &=&
-{1\over 2} \int _0 ^\infty {dT \over T} 
\nonumber\\[3mm]
 &\times&
\int {\cal D} x {\cal D}\psi
\, \exp 
\left\{ -\int _{\epsilon} ^T d\tau \, \left ( {1\over 2} \dot x ^\mu \dot x ^\mu + {1\over
2} \psi ^\mu \dot \psi ^\mu -{1\over 2} J_\Psi ^2  \right )\right\} 
\nonumber \\[3mm]
 &\times &  {\rm Tr }\,
{\cal P}\exp \left\{   i\int _0 ^T d\tau
\,  \left (A_\mu ^a \dot x^\mu -\frac{1}{2} \psi ^\mu F_{\mu \nu} ^a \psi ^\nu
\right ) T^a _r \right\}  \, ,
\eea
where 
the functional integral runs over all closed paths $x_\mu (\tau )$, 
$$
x_\mu (0) = x_\mu (T )\,,
$$
$A_\mu (x)$ is a fixed gauge background, and 
$$ T_r^a = T_{\rm adjoint}^a\,,  \,\,\, T_{\rm AS}^a\,,  \,\,\, T_{\rm S}^a$$
are the generators for the adjoint, two--index antisymmetric and
two--index  symmetric representations, respectively.
Moreover,  $\psi^\mu (\tau )$ are superpartners to
$x_\mu (\tau )$; they occur due to the fact that we are dealing with spin $1/2$ matter.

{} Eq.~(\ref{wlineint}) can be written symbolically as:
\beq
\Gamma _r [A, J_\Psi] =  
\sum_{\alpha} C_{\alpha} (J_\Psi) W_r^{\alpha}(A_{\mu}) \, ,
\label{indep}
\eeq
where the summation symbol also stands for the functional integrals appearing in (\ref{wlineint}).
The expansion
coefficients $C_{\alpha}(J_\Psi)$ depend, in general, on the representation
 $r$ through the sources $J_\Psi$. However, for the case at hand, the sources $J_\Psi$ can be matched in the three theories in such a way that, to leading order in $1/N$, the $C_{\alpha}(J_\Psi)$ become representation- {\it independent}. Examples of such a matching will be given at the end of this paper, also for the case of more than one flavor. With this in mind, we shall  assume hereafter that representation dependence resides entirely
in the (super) Wilson factors $W_r^{\alpha}(A_{\mu}) $.
Inserting the above result in (\ref{fint}) we arrive at
\beq
\label{expW}
e^{-{\cal W}_r(J_{\rm YM},\,J_\Psi)} =\left\langle e^{\sum_{\alpha} C_{\alpha}(J_\Psi) W_r^{\alpha}(A_{\mu}) }
\right\rangle  \, ,
\eeq
where the angle  brackets stand for the remaining functional integral (average) over the gauge field in the presence of a generic gluon source $J_{\rm YM}$.

\begin{figure}
\epsfxsize=6cm
\centerline{\epsfbox{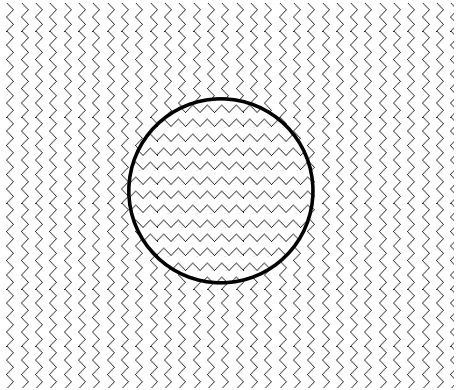}}
\caption{One fermion loop (i.e. $\log \, \det \left( i\not\! \partial + \not\!\!  A ^a
\, T^a _r + J_\Psi\right)$) in the gluon field background (shown as shaded areas).
The gluon fields ``inside" and ``outside"  the loop
do not communicate with each other at $N\to\infty$.
This is indicated by distinct shadings.
Averaging over the gluon field inside the loop is
independent of averaging outside. Topologically, of course, the distinction between inside and outside
is immaterial}
\label{Mone}
\end{figure}

\begin{figure}
\epsfxsize=9cm
\centerline{\epsfbox{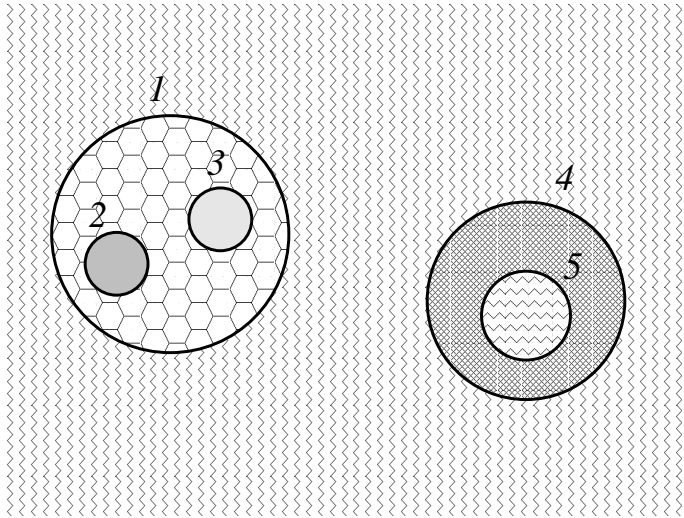}}
\caption{Example of fermion multiloops in the gluon field background,
at $N\to\infty$.
The background field outside loops 1  and 4 is the same.
The background field inside loop 1 and outside loop 2 and 3  is the same.
The background field inside loop 4 and outside loop 5  is the same.}
\label{Mtwo}
\end{figure}

As usual, taking the logarithm of both sides of (\ref{expW})  picks up the connected contributions from the expansion of the right-hand side,  
\bea
\label{W}
-{\cal W}_r(J_{\rm YM},\,J_\Psi) &=&\sum_n\,\,\,\,
 \sum_{\alpha_1, \alpha_2, \dots \alpha_n } \frac{C_{\alpha_1} \dots C_{\alpha_n}}{n!} \nonumber \\[4mm]
 &\times& \left\langle W_r^{\alpha_1}(A_{\mu})\,\,   W_r^{\alpha_2}(A_{\mu}) \,  \dots \, W_r^{\alpha_n}(A_{\mu})  \right\rangle_{c} \, ,
\eea
where the subscript $c$ stands for connected. In fact a subtlety, representing the
main thrust of this paper, is related to the issue of
``connectedness" which, in turn, is related to the
process of  averaging over the gluon field the multi-Wilson-loop operators appearing in Eq.~(\ref{W}).
In Ref.~\cite{Armoni:2003gp,Armoni:2004uu}
we dealt with a single loop (Fig.~\ref{Mone}),
now we will carefully treat multiloop averaging (see Fig.~\ref{Mtwo}
which displays, as a particular example, five loops).

We will compare two cases: $r=$adjoint and $r=$two--index antisymmetric.
The dimensions of the corresponding representations are
$$N^2-1\quad\mbox{and}\quad N(N-1)/2\,,$$ respectively. Note, however, that
the adjoint fermions are taken to be Majorana, while the two--index antisymmetric ones
are Dirac.  As a consequence, for $r=$adjoint,
Eq.~(\ref{wlineint}) has to be multiplied by $\frac{1}{2}$.
 Let us also note, in passing, that the  dimension of the two--index
symmetric representation is $N(N+1)/2$.

Our statement now is as follows:
{\sl As $N \rightarrow \infty$ each term in Eq.~(\ref{W}) for $r=$two--index antisymmetric has a corresponding term, with exactly the same value, for $r=$adjoint.} The proof is based on well-known trace identities. 

\vspace{1mm}

\begin{figure}
\epsfxsize=7cm
\centerline{\epsfbox{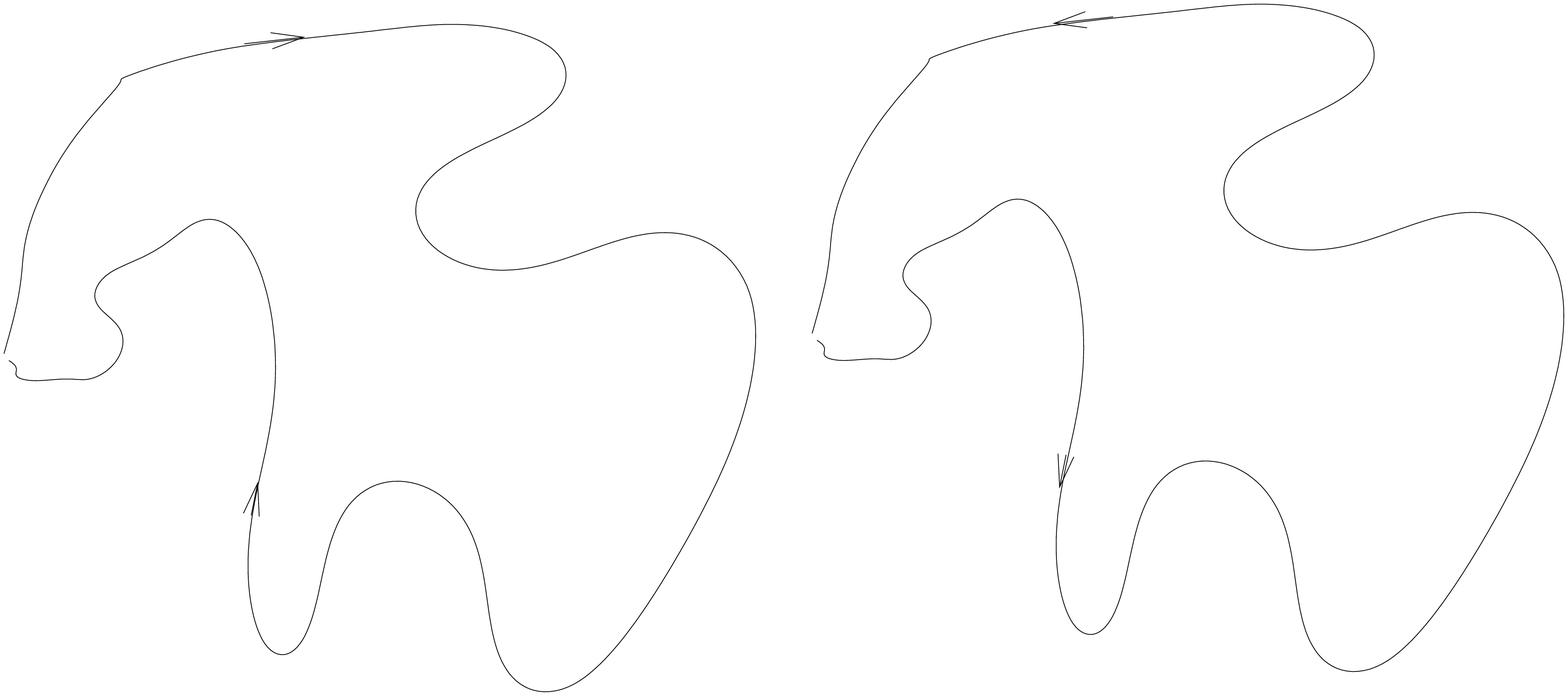}}
\caption{Two opposite-orientation contours in the sum (\ref{wlineint}).}
\label{Mthree}
\end{figure}

Since  $W_r^{\alpha}(A_{\mu})$, for a given
loop and  given $A_{\mu}$, is just the trace of one concrete SU$(N)$ group element, written in the representation $r$, the following relations hold (see, for example, \cite{Gross:1998gk}):
\bea
\label{traceid}
  W_{\rm S} &=&  {1\over 2} \left( ({\rm Tr}\, U)^2 + {\rm Tr}\,
U^2 \right)\, + (U\, \rightarrow U^\dagger) \label{halfS}
, \\[2mm]
W_{\rm AS} &=&  {1\over 2} \left( ({\rm Tr}\, U)^2 - {\rm Tr}\,
U^2 \right)\, +( U\, \rightarrow U^\dagger) \label{halfAS}
, \\[3mm]
  W_{\rm adjoint} &=& {\rm Tr}\, U\, {\rm Tr}\, U^\dagger -1 +
( U\, \rightarrow U^\dagger ) = 2  \left ({\rm Tr}\, U\, {\rm Tr}\, U^\dagger -1\right) \, ,
\label{ADJ}
\eea
where  $U$ (resp. $U^\dagger$) represents the same group element in the  
 {\em fundamental} (resp.  {\em antifundamental}) representation of $SU(N)$. 
 An important point  here is the occurrence of
the $( U\, \rightarrow U^\dagger) $ terms in Eqs.~(\ref{halfS}), (\ref{halfAS}).
The origin of these terms, whose presence is very natural since, for Dirac fermions, a representation and its complex conjugate are equivalent, can be explained as follows. 
 For each given oriented contour $x^\mu (\tau )$
in (\ref{wlineint}) there exists also the same contour with the opposite
orientation, see  Fig.~(\ref{Mthree}). We can therefore group 
Wilson loops pairwise (In QED this contour ``pairing" is responsible for the
Furry theorem.) and thus obtain the additional complex conjugate terms. For the real representations, such as the adjoint, this gives simply a  factor 2, as on the right-hand side of
Eq.~(\ref{ADJ}),  which cancels however in our case against the factor $\frac{1}{2}$ due to the Majorana condition.
 
\begin{figure}
\epsfxsize=10cm
\centerline{\epsfbox{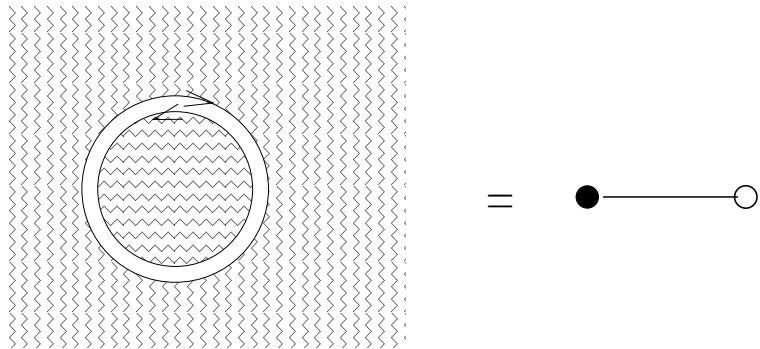}}
\caption{The 't Hooft double-line representation for $\langle W_{\rm adjoint} \rangle$. On the right we display a convenient graphic shorthand notation that we suggest
to use in this problem. The black circle corresponds to
$\langle {\rm Tr}\, U \rangle$, the white circle to $\langle {\rm Tr}\, U^\dagger \rangle$, while the segment connecting them indicates that  both circles originate from one and the same fermion loop, see Fig.~\ref{Mone}.}
\label{Mfour}
\end{figure}

Consider now, as the simplest example, the term with $n=1$ in (\ref{W}). At large $N$, after integrating over the gauge field, the  terms
$ {\rm Tr}\left( U^2\right)$ and $1$ are subleading 
in $1/N$ with respect to the terms of ${\cal O}(N^2)$ $({\rm Tr}\, U)^2$ or ${\rm Tr}\, U\, {\rm Tr}\, U^\dagger$. Furthermore, $\langle {\rm Tr}\, U \rangle\,  = \langle {\rm Tr}\, U^\dagger \rangle$. It then follows immediately that, at $N \rightarrow \infty$:
\beq
\label{simplest}
\frac{1}{2} \langle W_{\rm adjoint} \rangle \rightarrow  \langle W_{\rm S} \rangle \rightarrow  \langle W_{\rm AS} \rangle
\rightarrow  \langle  {\rm Tr}\, U  \rangle^2\,.
\eeq
Note that this common leading contribution is connected, in spite of the fact that, when written in terms of  Wilson loops with $r = $fundamental
(the last step in Eq.~(\ref{simplest})), it looks disconnected. 
It is instructive to graphically illustrate Eq.~(\ref{simplest}).
To this end we redraw Fig.~\ref{Mone} using the 't Hooft double-line notation, as shown in  Fig.~\ref{Mfour}. 

It is easy to show that, also  for higher-order  terms in Eq.~(\ref{W}), we can  drop the subleading contributions in  Eqs.~(\ref{traceid})--(\ref{ADJ}),
namely, $ {\rm Tr}\left( U^2\right)$ and $1$,
so that, hereafter, we will deal with the  large-$N$ limit,
\bea
W_{\rm S} &=& W_{\rm AS} =  {1\over 2} \left( ({\rm Tr}\, U)^2 + ({\rm Tr}\, U^{\dagger})^2\right) , 
\nonumber \\[2mm]
\frac{1}{2} W_{\rm adjoint} &=&   {\rm Tr}\, U\, {\rm Tr}\, U^\dagger  \, .
\label{traceidp}
\eea

Equations (\ref{traceidp}) suggest a convenient graphic 
notation (see again Figs.~\ref{Mone} and ~\ref{Mfour}). Associate with every
$W_{\rm adjoint}$ a black and a white  circle (related to ${\rm Tr}\, U$ and to ${\rm Tr}\, U^\dagger $, respectively)  connected by a short segment (just to show that it represents a {\it single} Wilson loop) and to either 
$W_{\rm S}$ or $W_{\rm AS}$ a similar drawing with two whites or two black circles (with a factor $\frac{1}{2}$ each). It is easy to see that, as 
$N\rightarrow \infty$, the leading diagrams for a generic contribution
of the form $\langle W_1 W_2 ... W_p\rangle _c$
in  Eq.~(\ref{W}) 
are given, in the above notation,  by a connected tree where $p$  segments are  joined through ``$i$-vertices",
i.e. vertices that couple any number $i$ ($i = 1, 2, \dots$) of dots. By using trivial properties of tree diagrams, and the fact that an $i$-vertex
gives a contribution ${\cal O}(N^{(2-i)})$, we arrive immediately to the conclusion that all tree diagrams are  ${\cal O}(N^2)$ while all loops are suppressed. It is amusing to notice that this 
large-$N$ counting resembles closely the one  of closed-string amplitudes if one associates with each
${\rm Tr}\, U$ or ${\rm Tr}\, U^\dagger $ a closed string and with each shaded region a tree-level
 vertex among the closed-strings that define that region's boundary.

\begin{figure}
\epsfxsize=6.5cm
\centerline{\epsfbox{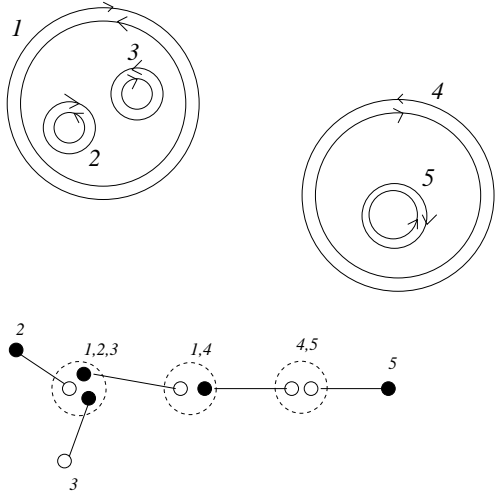}}
\caption{A particular connected  contribution to an expectation value
$\langle W_1 W_2 W_3 W_4 W_5\rangle _c$ in \None SYM theory.
Dashed circles indicate averaging over a connected background field,
for instance, the external lines of loop 1 and 
loop 4 are averaged over one and the same gluon field.}
\label{W1}
\end{figure}

We recall once more  that the subscript $c$ stands for
connected. In order to ease understanding of this point
we show, in Fig.~\ref{W1}, one of  contributions in the parent theory
(five fermion loops, see Fig.~\ref{Mtwo}) in the shorthand notation
introduced in Fig.~\ref{Mfour}.
\begin{figure}
\epsfxsize=6.5cm
\centerline{\epsfbox{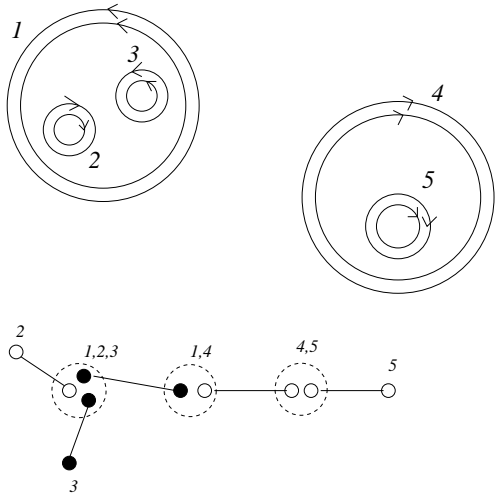}}
\caption{The same as in Fig.~\ref{W1}
after interchange ``black circle $\leftrightarrow$ white circle"
in vertices 5,  $(1,4)$, $2$ and $3$.}
\label{W2}
\end{figure}
This figure represents a certain large-$N$ connected correlator of five
Wilson loops in the parent theory, namely
 \bea
\langle W_1 W_2 W_3 W_4 W_5\rangle_c
&\longrightarrow&  \langle {\rm Tr\,} U_2 \rangle \,
\langle {\rm Tr\,} U_3^\dagger \rangle\,
\langle {\rm Tr\,} U_2^\dagger \, {\rm Tr\,} U_3\, {\rm Tr\,} U_1 \rangle
\nonumber\\[2mm]
&\times&
\langle {\rm Tr\,} U_1^\dagger\,  {\rm Tr\,} U_4 \rangle\,
 \langle {\rm Tr\,} U_4^\dagger\,{\rm Tr\,} U_5^\dagger  \rangle\,
\langle  {\rm Tr\,} U_5  \rangle\,.
\label{inson}
\eea

A similar contribution in the daughter theory (see Fig. 6) is
 \bea
\langle W_1 W_2 W_3 W_4 W_5\rangle_c
&\longrightarrow&  \langle {\rm Tr\,} U_2 ^\dagger \rangle \,
\langle {\rm Tr\,} U_3 \rangle\,
\langle {\rm Tr\,} U_2^\dagger \, {\rm Tr\,} U_3\, {\rm Tr\,} U_1 \rangle
\nonumber\\[2mm]
&\times&
\langle {\rm Tr\,} U_1 \,  {\rm Tr\,} U_4 ^\dagger \rangle\,
 \langle {\rm Tr\,} U_4^\dagger\,{\rm Tr\,} U_5^\dagger  \rangle\,
\langle  {\rm Tr\,} U_5 ^\dagger  \rangle\,.
\label{insond}
\eea
To complete the proof
of  the parent-daughter planar equivalence we now show  that, to every such tree diagram in the adjoint theory, one can associate a corresponding tree
shorthand diagram of the S or  AS  theory, having  exactly the same value.
To see that this is the case one can  interchange white and black circles   at every other vertex  along the tree
as shown in Fig.~\ref{W2}. After doing so we arrive at a
five fermion loop contribution in the
daughter (A or S) theory. 
This operation obviously transforms a generic graph of the adjoint theory into a corresponding graph of the $S$ or $AS$ theory. 
The fact that one can perform the interchange ``black circle 
$\leftrightarrow$ white circle" at any vertex separately, is rather obvious. 
Take, for instance, the vertex (1,4) in Figs.~\ref{W1} and \ref{W2}.
The above interchange is nothing but the use of an obvious equality
\beq
\langle {\rm Tr\,} U_1^\dagger\,  {\rm Tr\,} U_4 \rangle
=
\langle {\rm Tr\,} U_1\,  {\rm Tr\,} U_4^\dagger \rangle\,,
\eeq
which is a straightforward generalization of
the equality  $\langle {\rm Tr}\, U \rangle\,  = \langle {\rm Tr}\, U^\dagger \rangle$. It is easy to show that the above procedure is  biunivocal i.e. it associates to every graph of the parent  theory a graph of the daughter theory, and 
{\em vice versa}.

Other  one-to-one transformations of the 
graphs of the parent theory into those of the daughter theories 
are also possible.
For instance, one can use the fact that $W_{\rm adjoint}$
is real even before averaging over the gluon field, see the second line in Eq.~(\ref{traceidp}). This means, in essence, that
the loops of the parent theory are unoriented (the consequence of the
reality of the adjoint representation). In addition, it
is not necessary  to isolate and ``pair" together, from the very beginning, 
contours of the opposite orientation, as shown in Fig.~\ref{Mthree}.
One can let the sum run over all contours independently.
This will lead to untangling the terms
$({\rm Tr}\, U)^2$ and $({\rm Tr}\, U^\dagger)^2$
in the first line in Eq.~(\ref{traceidp}).
They will appear as separate contributions. 
And, nevertheless, each of these separate
contributions will have an equal counterpart in the
parent theory. 

To conclude, we proved a non-perturbative equivalence between the
partition function of \None SYM theory and ``orientifold field
theory''. The equivalence holds also in the presence of certain external currents.
While we do not provide an exact detailed dictionary of the ``common
sector'' of the two theories, it is clear from our proof that
correlation functions that involve powers of ${\rm Tr}\, F^2$ as well as
${\rm Tr}\, F \tilde F$ match at large $N$. The bifermion operators
$\bar \Psi\Psi$ and $\bar \Psi\gamma_5\Psi$ are also in the common sector,
and so is the axial current $\bar \Psi\gamma_\mu \gamma_5\Psi$.
The vector current $\bar \Psi \gamma_\mu \Psi$ and the tensor
operator $\bar \Psi \sigma_{\mu\nu} \Psi$ do not belong to the common sector, however.
>From our proof it follows that the
bosonic hadron spectra as well as the domain wall spectra (mass and charge) 
are the same in the two theories. In general, every operator in the
parent theory that survives the orientifold projection belongs to the
common sector.\footnote{In the gauge/string correspondence these 
operators couple to closed string modes which
are common to type IIB and type 0'B string theories.} 

Finally, we would like to briefly discuss a
rather obvious generalization which had been called \cite{Armoni:2003jk}
flavor proliferation. Our proof of planar equivalence can be readily
generalized to the case of many flavors:   

{\sl SU($N$) Yang--Mills theory with $N_f$ Majorana
fermions in the adjoint representation (non-supersymmetric if $N_f>1$)
is equivalent, in the common sector, to   SU($N$) Yang--Mills theory
with $N_f$ Dirac fermions in the two--index
antisymmetric representation}\,\footnote{In general, the $m \rightarrow
0$ and   $1/N \rightarrow 0$ limits need not (and do not) commute. 
We consider here planar equivalence, namely we take the 
limit $1/N \rightarrow 0$ first.}. The common sector includes
all operators built of gluon fields and covariant derivatives,
and a subset of bifermion operators.

A few explanatory remarks are in order here regarding the
determination of the common sector in the multiflavor case. 
In selecting bifermion operators that belong
to the common sector (i.e. the set of sources $J_\Psi$)
one should exercise caution. The pattern of flavor symmetry
in these two theories are drastically different. In the parent theory with
the Majorana fermions, the global flavor
symmetry is SU$(N_f)$; it is believed to be spontaneously broken down to
SO$(N_f)$, see e.g. Ref.~\cite{Kogan:1984nb}, while in the daughter theory
the pattern is the same as in QCD, namely
\beq
{\rm SU}(N_f)_L\times {\rm SU}(N_f)_R
\to {\rm SU}(N_f)_V\,.
\label{tsaqcd}
\eeq
This is the reason why many fermion bilinears
do not belong to the common sector.

Operationally, it can be defined as follows.
Start from the parent theory with $N_f$ Majorana flavors in 
the adjoint. Write all possible bilinears which do not vanish 
by symmetry. Perform orientifoldization and find the projection of the 
above set to the daughter. Call this `` common class"  C.
Alternatively, one can also start from the daughter theory.
Write all possible bilinears. Examine which ones of them  can 
be elevated to the parent theory. These should define the same class C. 

There is a large number of  fermion bilinears which do not lie in 
C in the parent theory, 
and the same is true for  the daughter theory.
These do not belong to the common sector.
 
For fermion bilinears with {\em no derivatives} one can
readily present a complete catalogue. 
With respect to the Lorentz symmetry they form the following
representations: 
$$
(0,0)\,,\qquad \left(\frac{1}{2}, \frac{1}{2}\right)\,,\qquad (0,1)+ (1,0)\,.
$$
The $(0,0)$ operators in the parent theory are of the type
\beq
{\rm Tr}\,\lambda_\alpha^{ f} \lambda^{\alpha\, g}
\label{tys}
\eeq
where $f$ and $g$ are the flavor indices. To have a non-vanishing operator, 
we must symmetrize with respect to $f,g$. Altogether we get
$N_f(N_f+1)/2$ operators of the type (\ref{tys}) plus $N_f(N_f+1)/2$
complex conjugate operators. 
Let us denote the orientifold projection of $\lambda_i^{j\,,f}$
as
\beq
\lambda_i^{j\,,f} \to \{\,  \chi_{[ij]}^f\,,\,\, \eta^{[ij]}_f\, \}
\label{orre}
\eeq
where $i,j$ are the color indices and $\eta\,,\chi$ are chiral
(left-handed) spinors of the daughter theory. Each pair $\chi\,,\bar\eta$
forms one Dirac flavor. It is clear that (\ref{tys})
projects onto $\chi^f\eta_g + \chi^g\eta_f$.
In Dirac notation the projection is  onto
$$
 \bar\Psi_f\frac{1+\gamma_5}{2} \Psi^g\,,\qquad f,g\mbox{-symmetrized}\,,
$$
 and similarly for the complex-conjugate bilinears.
The $(0,1)+ (1,0)$ operators in the parent theory are of the type
\beq
{\rm Tr}\,\lambda_{\{\alpha}^{[ f} \lambda_{\beta\}}^{ g]}\,,
\label{tysp}
\eeq
with symmetrized $\alpha\,,\, \beta$ and antysymetrized $f,g$.
There are $N_f(N_f-1)/2$ operators (\ref{tysp}) and the same amount of complex conjugate operators. They project onto
$$
\bar\Psi_{f}\sigma_{\mu\nu}\Psi^{g}\,,\qquad f,g\mbox{-antisymmetrized},
$$
in the daughter theory.

Finally, the operators $(1/2,1/2)$ are currents. Their classification is discussed in sufficient detail in Ref.~\cite{Kogan:1984nb}. The total number of currents in the parent theory is $N_f^2$ (including one anomalous current), while
the total number of currents in the daugther  theory is $2N_f^2$
(namely, $N_f^2$ vector  and $N_f^2$ axial currents). It is clear that a half
of the daughter theory currents have no projection onto the parent one.

What currents can be projected? 
To answer this question we can use, again, the basic projection
(\ref{orre}). The $N_f^2$ currents of the parent theory are
\beq
\bar\lambda_{\dot\alpha\,, g}\, \lambda_\beta^f\,,\qquad f,g = 1,2,..., N_f\,.
\eeq
They are projected as
\beq
\bar\chi_{\dot\alpha\,, g}\, \chi_\beta^f
+ \bar\eta_{\dot\alpha}^g \, \eta_{\beta\,, f}\,.
\label{c1}
\eeq
The daughter theory has $2N_f^2$ currents,
\beq
\bar\chi_{\dot\alpha\,, g}\, \chi_\beta^f\,,\quad\mbox{and}\quad \bar\eta_{\dot\alpha}^g \, \eta_{\beta\,, f}\,.
\label{c2}
\eeq
Comparing Eqs. (\ref{c1}) and (\ref{c2})
we conclude that the minus combination of the currents in (\ref{c2})
does not make it to the common sector.

The analysis becomes even  easier
 if we use the Majorana rather than Weyl's representation
of the adjoint spinors in the parent theory.
In this case  the nonvanishing
$(1/2,1/2)$ operators in the parent theory are
\beq
\bar\lambda ^{[f}\gamma_\mu\lambda^{g]}\,,\qquad
\bar\lambda ^{\{f}\gamma_\mu\gamma_5 \lambda^{g\}}\,.
\label{mity}
\eeq
where curly and square brackets denote symmetrization and antisymmetrization,
respectively.
The total number of the currents (\ref{mity}) is $N_f^2$.
Performing orientifoldization (i.e. replacing $\lambda \to \Psi$ and $\bar\lambda \to \bar\Psi$)
we get $N_f^2$ currents of the daughter theory which belong 
to the common sector.\footnote{For instance, 
in this way it is easy to check that ``extra" Goldstone bosons
that exist in the daughter theory but are absent from the
parent one \cite{Armoni:2003jk} do not belong to the common sector.
They can be produced only in pairs. This contribution
is subleading in $1/N$.} Their charges generate an unconventional ${\rm SU}(N_f)$ subgroup of 
${\rm SU}(N_f)_L\times {\rm SU}(N_f)_R$, containing both vector and axial transformations.

If we allow for no-derivative bifermion operators {\em with} gluon fields
included,
the set of allowed operators expands dramatically . We will 
make no attempt at a complete classification in this case. Let us give just
one example.
With a single  insertion of the gluon field, one
can build combinations
$$
{\rm Tr}\,  \lambda_\rho^f \, F_{\alpha\beta}\,  \lambda_\gamma^g
\,\, \varepsilon^{\rho \alpha}
$$
with all possible symmetry patterns for $\gamma \,,\,\beta$ and $f,g$.

Inclusion of derivatives leads to a further enlargement of the common sector.
The issue of a complete classification in this case is left  for future work.

\vspace{1cm}

{\bf Acknowledgments}
We would like to thank L. Alvarez-Gaum\'{e}, R.~Casalbuoni, L. Del-Debbio, 
P. Di~Vecchia, M. L\"{u}scher, R. Musto and A. Ritz for very useful discussions. The work  of M.S. was supported in part by DOE grant DE-FG02-94ER408. 
A.A. is supported by the PPARC advanced fellowship.

\end{document}